\documentstyle[a4,12pt,amssymb,float,psfig,times]{article}
\newcommand{\be}{\begin{equation}}
\newcommand{\ee}{\end{equation}}
\newcommand{\bea}{\begin{eqnarray}}
\newcommand{\eea}{\end{eqnarray}}

\def\p1{\pi_1}
\def\b{\beta}
\def\l{\lambda}
\def\f{\phi}
\def\r{\rho}

\def\g{\gamma}
\def\pmas{\partial_+}
\def\pmen{\partial_-}

\begin{document}

\begin{titlepage}
\vspace*{\stretch{0}}
\begin{flushright}
{\tt FTUV/99-42\\
     IFIC/99-44\\
     SU-ITP/99-29\\
     hep-th/9906187}
\end{flushright}

\vspace*{0.5cm}

\begin{center}
{\Large\bf Integrable models and degenerate horizons\\ 
 in two-dimensional gravity}
\\[0.5cm]
J. Cruz$^1$\footnote{E-mail address: cruz@lie.uv.es}, A. Fabbri$^2$
\footnote{E-mail
address: afabbri1@leland.stanford.edu}, D. J. Navarro$^{1,2}$\footnote{E-mail
address: dnavarro@ific.uv.es} and J. Navarro-Salas$^1$
\footnote{E-mail address:
jnavarro@lie.uv.es}.
\\ 
{\footnotesize
       1) Departamento de F\'{\i}sica Te\'orica and
       IFIC, Centro Mixto Universidad de Valencia-CSIC.\\
       Facultad de F\'{\i}sica, Universidad de Valencia,
       Burjassot-46100, Valencia, Spain.\\
       2) Department of Physics, Stanford University, Stanford, CA,
94305-4060,
       USA.}
\end{center}
\bigskip
\begin{abstract}
We analyse an integrable model of two-dimensional gravity which can be reduced 
to a pair of Liouville fields in conformal gauge. Its general solution 
represents a pair of ``mirror'' black holes with the same temperature. The 
ground state is a degenerate constant dilaton configuration similar to the
Nariai solution of the Schwarzschild-de Sitter case. The existence of 
$\phi=const.$ solutions and their relation with the solution given by the 
2D Birkhoff's theorem is then investigated in a more general context. We also 
point out some interesting features of the semiclassical theory of our model
and the similarity with the behaviour of AdS$_2$ black holes.

\end{abstract}
\vspace*{\stretch{1}}
\begin{flushleft}
PACS number(s): 04.60.Kz, 04.60.Ds
\end{flushleft}
\vspace*{\stretch{0}}
\end{titlepage}
\newpage

\section{Introduction}
The existence of exactly solvable models of gravity in two dimensions
\cite{bc} provides a rich arena for the study of quantum aspects of black
holes. These two-dimensional black holes, in addition to their own interest,
can describe particular regimes of higher-dimensional black holes. The CGHS
model \cite{cghs} describes low-energy excitations of extremal (magnetic)
string black holes in four dimensions. AdS$_2$ black holes arise in the
near-horizon limits of extremal or near-extremal Reissner-N\"ordstrom black holes
\cite{c,mms}. By dimensional reduction, spherically symmetric gravity can
also be described in terms of an effective two-dimensional model.\\

The aim of this paper is to analyse a general family of integrable models
\cite{f} which can recover all known solvable models (CGHS \cite{cghs},
Jackiw-Teitelboim \cite{jt} and exponential (Liouville) \cite{cfn} models) in
some particular limits. The equations of motion of the models, in conformal
gauge, are equivalent to those of a pair of Liouville fields for linear
combinations of the conformal factor and the dilaton field. These properties
will be briefly reviewed in section 2. In section 3 we investigate the
properties of the classical solutions showing that, in the absence of matter
fields, they represent a pair of eternal black holes. 
In section 4 we shall focus 
in one particular model (with a potential of the form $V(\f)=2\sinh \b
\f$) which allows a degenerate solution having a constant value for the
dilaton and a two-dimensional de Sitter (or anti-de Sitter, depending on the sign of the
constant $\b$)
geometry. The situation is similar
to that encountered in the Schwarzschild-de Sitter case where the
degenerate case of the Nariai metric \cite{n} is also described by a constant
dilaton (i.e. the radial coordinate). In section 5 we shall analyse the
existence of such dilaton-constant solutions in a more general setting. We
will show in a simple way that these configurations are possible for the
zeros of the potential, after removing the kinetic term of the
two-dimensional dilaton-gravity theory, and are always accompanied by a
constant curvature geometry. Furthemore, they are always connected with the 
presence of degenerate horizons in the theory.  
Finally, in section 6 we make some comments on the semiclassical behaviour 
of our solutions and show 
interesting similarities with the behaviour of AdS$_2$ black holes.

\section{Integrability of 2D dilaton gravity models}
Let us consider the general functional action describing a 2D dilaton gravity
model coupled to $N$ 2D massless and minimal scalar fields
\be
\label{action2D} S=\frac{1}{2\pi} \int d^2x \sqrt{-g} \left( R\phi + 4\l^2
V(\f) - \frac{1}{2} \sum_{i=1}^N (\nabla f_i)^2 \right) \, , \ee where
$V(\f)$ is an arbitrary function of the dilaton field and $f_i$ are the scalar
matter fields. The above expression represents a generic model because one can
get rid of the kinetic term of the dilaton by a conformal reparametrization
of the fields and bring the action into the form (\ref{action2D})
\cite{lmgk}. In conformal gauge $ds^2=-e^{2\r}dx^+dx^-$, the equations of
motion derived from the action (\ref{action2D}) are 
\bea \label{eq1}
2\pmas\pmen\r + \l^2 V^{\prime}(\f) e^{2\r} & = & 0 \, , \\ \label{eq2}
\pmas \pmen \f + \l^2 V(\f) e^{2\r} & = & 0 \, , \\ \label{eq3} \pmas \pmen
f_i & = & 0 \, , \\ \label{eq4} -\partial_{\pm}^2 \f + 2\partial_{\pm} \f
\partial_{\pm} \r -\frac{1}{2} \sum_{i=1}^N (\partial_{\pm}f_i)^2 & = & 0 \,
. \eea 
By introducing an arbitrary parameter $\b$ we can rewrite the above
equations of motion (\ref{eq1}), (\ref{eq2}) in the form \bea
\pmas\pmen(2\r+\b\f) + \l^2 e^{2\r} \left( \b V(\f) + \frac{dV(\f)}{d\f}
\right) &=& 0 \, , \\ \pmas\pmen(2\r-\b\f) - \l^2 e^{2\r} \left( \b V(\f) -
\frac{dV(\f)}{d\f} \right) &=& 0 \, . \eea One way to ensure the
integrability of the above equations is to reduce them to a pair of Liouville
equations \cite{f}.
 The most general potential satisfying this requirement is
\be
\label{2exp} V(\f) = \g_+ e^{\b\f} + \g_- e^{-\b\f} \, , \ee so that the
corresponding equations of motion are a pair of Liouville equations
\be
\label{leq} \pmas\pmen(2\r\pm\b\f) \pm 2\g_{\pm} \b \l^2 e^{2\r\pm\b\f} = 0
\, . \ee 
This potential includes all known integrable models.
That is, for $\g_{\pm}=\frac{1}{2}$ and $\b \rightarrow 0$ the CGHS model;
$\g_+=-\g_-= \frac{1}{2\b}$ and $\b \rightarrow 0$ the Jackiw-Teitelboim
theory and for $\g_+=1$, $\g_-=0$ the exponential (Liouville) model \cite{cinn}.\\

The general solution to the equations (\ref{leq}) can be written in terms of
four arbitrary chiral functions $A_{\pm}(x^{\pm})$, $a_{\pm}(x^{\pm})$ \bea
\label{ls1} 2\r+\b\f &=& \ln \frac{\pmas A_+ \pmen A_-}{(1+\g_+ \b \l^2
A_+A_-)^2} \, , \\ \label{ls2} 2\r-\b\f &=& \ln \frac{\pmas a_+ \pmen
a_-}{(1-\g_- \b \l^2 a_+a_-)^2} \,  \eea and allows to recover the general
solution of the limiting models. The solution for the exponential model is
immediately recovered making $\g_+=1$ and $\g_-=0$ in (\ref{ls1}),
(\ref{ls2}). In the Jackiw-Teitelboim theory ($\g_+=-\g_-=\frac{1}{2\b}$) we
have to redefine the functions $a_{\pm}$ introducing a new pair
$\tilde{a}_{\pm}$, $a_{\pm}=A_{\pm}+\b \tilde{a}_{\pm}$. Afterwards we
realize the $\b \rightarrow 0$ limit and then we get \bea \r &=& \phantom{+}
\frac{1}{2} \ln \frac{\pmas A_+ \pmen A_-}{(1+\frac{\l^2}{2} A_+ A_-)^2} \, ,
\\ \f &=& -\frac{1}{2} \left( \frac{\pmas \tilde{a}_+}{\pmas A_+} +
\frac{\pmen \tilde{a}_-}{\pmen A_-} \right) + \frac{\l^2}{2} \frac{A_+
\tilde{a}_- + A_- \tilde{a}_+}{1+\frac{\l^2}{2} A_+ A_-} \, , \eea as it was
found in \cite{cinn}. Finally we can also recover the solution for the CGHS
model ($\g_{\pm}=\frac{1}{2}$) in a similar way. Redefining $a_{\pm} =
A_{\pm} -2\b \int^{x^{\pm}} \hat{a}_{\pm} \partial_{\pm} A_{\pm}$ in
(\ref{ls1}), (\ref{ls2}) we can perform the $\b \rightarrow 0$ limit and we
get \bea \r &=& \frac{1}{2} \ln \pmas A_+ \pmen A_- \, , \\ \f &=& -\l^2 A_+
A_- + \hat{a}_+ + \hat{a}_- \, . \eea

The above mechanism provides a very simple picture on the origin of the
integrability of these models and suggests a particular analysis of the most
general integrable hyperbolic model (\ref{2exp}). The hidden reason of this
integrability can now be understood as all them are particular cases of a general
Liouville integrability of which the hyperbolic model is, in a sense, the maximal
one. The hyperbolic model is then the most complicated solvable model that we can study.

\section{Classical theory and eternal black hole solutions}
In this section we shall study the classical theory of the model (\ref{2exp})
and look for black hole solutions. The functional action is given by
\be
\label{2expaction} S=\frac{1}{2\pi} \int d^2x \sqrt{-g} \left( R\phi + 4\l^2
(\g_+ e^{\b\f} + \g_- e^{-\b\f}) - \frac{1}{2}\sum_{i=1}^N (\nabla f_i)^2 \right) \,  \ee
and we have to note that, although one of the three parameters $\l$, $\g_+$,
$\g_-$ is redundant, we shall maintain all of them in order to simplify the
equations.\\

The solutions to the unconstrained equations of motion of the above theory
are given by (\ref{ls1}) and (\ref{ls2}). Now, in terms of the $A_{\pm}$,
$a_{\pm}$ functions the constraint equations (\ref{eq4}) become
\be
T^{f}_{\pm\pm} = -\frac{1}{2\b} \left( \left\{ A_{\pm},x^{\pm} \right\}
-
\left\{ a_{\pm},x^{\pm} \right\} \right) \, , \ee where $\{ \; , \; \}$
denotes the Schwartzian derivative.\\

In the absence of matter fields and in an appropriate Kruskal-type gauge
$a_{\pm}=x^{\pm}$ the general solution is given by \bea \label{conf1} ds^2
&=& \frac{-dx^+dx^-}{(\frac{\l^2\b}{C}+\g_+ Cx^+x^-) (1-\g_-
\l^2 \b x^+x^-)} \, , \\ \label{conf2} e^{\b\f} &=& \frac{1-\g_- \l^2 \b
x^+x^-}{\frac{\l^2\b}{C}+\g_+ Cx^+x^-} \, , \eea where the parameter $C$ is
related with the conserved quantity $M$ (proportional to the ADM mass)
\be
\label{mass} M = \frac{1}{\b} \left( \frac{C}{\l^2 \b} \g_+ - \frac{\l^2
\b}{C} \g_-
-
\g_+ + \g_- \right) \, . \ee 
In a `pure' two-dimensional context and in order to study the full spacetime
structure of the solution we will place no restriction on the range of variation
of the field $\phi$. Of course, if our starting point were four dimensional 
the identification of $\phi$  
with the radius of the two-sphere $r$ would imply that only $\phi>0$ is allowed.
The curvature of the solution is
\be
\label{curvature} R = -4\l^2 \b \left( \g_+ \frac{1-\g_- \l^2 \b
x^+x^-}{\frac{\l^2\b}{C}+ \g_+ Cx^+x^-} - \g_- \frac{\frac{\l^2\b}{C}+\g_+
Cx^+x^-}{1-\g_- \l^2
\b
x^+x^-} \right) \,  \ee and there are two curvature singularities at \bea
\label{sing1} x^+x^- &=& \frac{-\l^2 \b}{\g_+ C^2} \, , \\ \label{sing2}
x^+x^- &=& \frac{1}{\g_- \l^2 \b} \, . \eea 
In order to avoid timelike
singularities we have two possibilities: $\b<0$, $\g_+>0$, $\g_-<0$, or
$\b>0$, $\g_+<0$, $\g_->0$. They are actually the same because the potential
(\ref{2exp}) is symmetric under the interchange of both cases. The Kruskal 
diagram is represented in Fig. I.
\begin{figure}[H]
\centerline{\psfig{figure=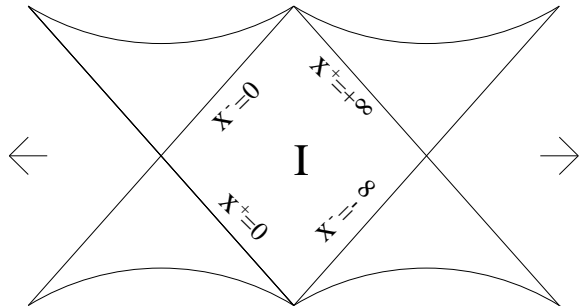,width=3.in,angle=-90}}
\caption{Kruskal diagram for the hyperbolic model}
\end{figure}
The horizons ($\partial_{\pm} e^{\b \f}=0$) are located at $x^{\pm}=0, \pm
\infty$ respectively. The Killing vector $\frac{\partial}{\partial t}$ is
timelike in the regions I and spacelike in the others. 
\\
Choosing $\b<0$ we can define $\g>0$ so that $-\frac{\g_+}{\g_-}=
\frac{1}{\g}$ and we are able to redefine the parameter $\l$ in order to
absorb the extra parameter. In this way, an hyperbolic model having eternal
black hole solutions is given by the following functional action
\be
\label{sinhaction} S=\frac{1}{2\pi} \int d^2x \sqrt{-g} \left( R\phi + 4\l^2
(e^{\b\f} - \g e^{-\b\f}) - \frac{1}{2} \sum_{i=1}^N (\nabla f_i)^2 \right)
\, . \ee
It is interesting to note that we do not lose generality by restricting
to the case  $\g=1$. In fact, even if we consider $\g\neq 1$ the  
redefinitions $e^{\b\f}\to e^{\b\f}\sqrt{\g}$ and 
$\lambda^2\sqrt{\g} \to \lambda^2$ recast the potential in the $\g=1$ form.
Moreover, the constant shift in the field $\f$ will produce an extra piece in the 
action proportional to $R$, but this, being just a boundary term, does not
affect the equations of motion. We then consider the following potential
\be
\label{sinh} V(\f)= 2\sinh \b \f \, . \ee Its geometry is given by the metric
\be
\label{sinhmetric} ds^2 = \frac{-dx^+dx^-}{(\frac{\l^2\b}{C}+Cx^+x^-)(1+\l^2
\b x^+x^-)} \, , \ee with dilaton function
\be
\label{sinhdilaton} e^{\b\f} = \frac{1+\l^2 \b
x^+x^-}{\frac{\l^2\b}{C}+Cx^+x^-} \, ,  \ee and $M$ and curvature read 
\bea \label{sinhmass} M &=& \frac{1}{\b} \left( \frac{C}{\l^2 \b} +
\frac{\l^2 \b}{C} - 2 \right) \, , \\ \label{sinhcurvature} R &=& -4\l^2 \b
\left(\frac{1+ \l^2 \b x^+x^-}{\frac{\l^2\b}{C}+Cx^+x^-} +
\frac{\frac{\l^2\b}{C}+Cx^+x^-}{1+\l^2 \b x^+x^-} \right) \, . \eea 
This
model is interesting due to the presence of a dilaton-constant
solution. The curvature has generically two singularities at points (\ref{sing1}),
(\ref{sing2}) ($\g_+=-\g_-=1$). However, in the limit $C \rightarrow \l^2 \b$ it
becomes regular and constant everywhere and the dilaton field
is constant $e^{\b \f} = 1$. 
The similarity of this solution with a known one in Einstein gravity will be 
explored in the next section.

\section{Degenerate horizon solutions and comparison with the Schwarzschild-de
Sitter case}
In this section we shall study the particular $\phi=0$
solution of the model (\ref{sinh}) because it has a special similarity
with the Nariai solution appearing in the Schwarz\-schild-de Sitter solution
\cite{bh,gp}. The Schwarzschild-de Sitter metric is the static spherically
symmetric solution of the Einstein equations with a cosmological constant
$\Lambda$. It is
\be
ds^2 = -\tilde{U}(r)dt^2 + \tilde{U}(r)^{-1} dr^2 + r^2 d\Omega^2 \, , \ee
where
\be
\tilde{U}(r) = 1 - \frac{2m}{r} - \frac{\Lambda}{3} r^2 \, . \ee For
$0<m<\frac{1}{3} \Lambda^{-\frac{1}{2}}$, $\tilde{U}(r)$ has two positive
roots corresponding to the black hole and cosmological horizons. But in the
limit $m \rightarrow \frac{1}{3} \Lambda^{-\frac{1}{2}}$ the two roots
coincide and the horizons apparently merge. In this degenerate case the
Schwarzschild coordinates become inappropriate since $\tilde{U}(r)
\rightarrow 0$ between the two horizons. 
According to Ginsparg and Perry \cite{gp} we can define new coordinates $\psi$ and
$\chi$
\be
t={1\over\epsilon\sqrt{\Lambda}}\psi,\qquad r={1\over\sqrt{\Lambda}}\left[
1-\epsilon\cos \chi-{1\over6}\epsilon^2\right]\>,
\ee
where
\be
9m^2\Lambda=1-3\epsilon^2
\ee
with the property that the new metric has a well-defined limit 
in the degenerate case $\epsilon\rightarrow 0$
\be
ds^2=-{1\over\Lambda}\left(\sin^2\chi d\psi^2-d\chi^2\right)+{1\over\Lambda}
d\Omega^2\>,
\ee
which turns out to be the Nariai solution.

A similar situation is found in the
model (\ref{sinh}). To see this feature we consider the static solution (we
call this the Schwarzschild gauge) that the 2D Birkhoff's theorem \cite{lmk}
provides for a generic model (\ref{action2D}). This solution is written as
\bea
\label{birk1} ds^2 &=& -(4J(\f)-4M)dt^2 + (4J(\f)-4M)^{-1}dr^2 \, , \\
\label{birk2} \f &=& \l r \, , \eea where
\be
\label{ADMmass} M = J(\f) -\frac{1}{4\l^2} (\nabla \f )^2 \,  \ee is a
diffeomorphism invariant parameter related with the ADM mass and
$J(\f)=\int_0^{\f} d\tilde{\f} V(\tilde{\f})$.\\
For the model (\ref{sinh}) we get the following metric
\be
\label{Smetric} ds^2 = -U(r) dt^2 + U(r)^{-1} dr^2 \,  , \ee where
\be
\label{lulo}
U(r) = \frac{8}{\b} (\cosh \l \b r -1) -4M \, . \ee If we consider $\b<0$,
$M\leq 0$ solutions, there are two horizons ($U(r_{\pm})=0$) located at
\be
r_{\pm} = \pm \frac{1}{\l\b} \mathrm{arcosh} (1+\frac{1}{2} \b M) \, , \ee
but in the limit $M \rightarrow 0$ the horizons become coincident
($r_{\pm}=0$) and $U(r) \rightarrow 0$ between them. There are two curvature
singularities at $r=\pm \infty$ since the curvature is
\be
R = -8\l^2 \b \cosh \l \b r \, . \ee We can interpret this solution as two
``mirror'' black holes located `at infinity' hidden by two horizons
$r_{\pm}$. The space-time between the horizons admits a timelike Killing
vector ($U(r)>0$), which becomes spacelike behind the horizons ($U(r)<0$). In
the limit $M \rightarrow 0$ the horizons coalesce and in this region $U(r)
\rightarrow 0$, $\f \rightarrow 0$, $R \rightarrow -8\l^2 \b$.
 In this limit the $(t,r)$ coordinates become inappropriate and we
need to perform a coordinate change.\\

If we define a parameter $C$ so that $M$ is written as (\ref{sinhmass}) the
$M \rightarrow 0$ limit is recovered in the $C \rightarrow \l^2 \b$ limit.
Thus let us try the following transformation \bea \label{gp1} -x^+x^- &=&
\frac{\frac{\l^2 \b}{C} - e^{-\l \b r}}{C-\l^2 \b e^{-\l \b r}} \, , \\
\label{gp2} -\frac{x^+}{x^-} &=& e^{-\l \b (\frac{\l^2 \b}{C} -\frac{C}{\l^2
\b})t} \, , \eea relating both Kruskal and Schwarzschild gauges as it brings
(\ref{Smetric}), (\ref{lulo}), (\ref{birk2}) into (\ref{sinhmetric}), (\ref{sinhdilaton})
respectively. This transformation is singular for the degenerate case $C=\l^2
\b$ ($M=0$) as the Ginsparg-Perry one for the 4D Schwarzschild-de Sitter
gravity \cite{gp}. We can actually see it as a perturbation around the point
$r=0$ where both horizons coincide. When $C=\l^2 \b$ there are no
singularities and the metric (\ref{sinhmetric}) turns into
\be
\label{ds} ds^2=\frac{-dx^+dx^-}{(1+\l^2\b x^+x^-)^2} \,  \ee and, finally,
the new transformation
\be
x^{\pm} = \frac{1}{\l\sqrt{-\b}} (\sinh \psi \pm \cosh \psi) \frac{\sin
\chi}{1+\cos \chi} \,  \ee brings it into the 2D-reduced part of the
Nariai solution \cite{n} with topology $H^2$
\be
\label{metric} ds^2 = \frac{-1}{4\l^2 \b} (-\sin^2 \chi d\psi^2 + d\chi^2) \,
. \ee Note that even though the transformation (\ref{gp1}), (\ref{gp2}) is
singular for the degenerate case, the coordinates $x^{\pm}$ remain
appropriate for this case too and the horizons' radii also remain
different. The true reason for which this transformation becomes singular in
the limit $C \rightarrow \l^2 \b$ is due to the fact that both Kruskal gauge (constant
dilaton) and Schwarzschild gauge (linear dilaton) solutions are not
diffeomorphism connected. They are indeed two different solutions
 and this motivates a revision of the 2D Birkhoff's theorem
which will be made in the next section.\\

To finish this section we shall consider the thermodynamics of this model.
Since the static Schwarzschild gauge (\ref{Smetric}) is not the appropriate
one to study the thermodynamics due to the degenerate limit we look for
another one starting from the conformal-Kruskal gauge (\ref{sinhmetric}).
This is possible since the model always admits a timelike Killing vector.
Thus let us introduce new static coordinates $y^{\pm}$ given by
\be
\label{strans-} \pm \omega x^{\pm} = e^{\pm \omega y^{\pm}} \, , \ee where
$\omega^2 = -C$. In terms of these coordinates the metric becomes
\be
ds^2 = \frac{-dt^2 + dy^2}{(1-\frac{\l^2 \b}{C})^2 + \frac{4 \l^2 \b}{C}
\cosh^2 \omega y} \, . \ee 
The metric is manifestly static in this form and
it is straightforward to find a new Schwarzschild-type gauge by means of the
new spacelike coordinates defined by
\be
\sigma = \frac{1}{\omega (1+\frac{\l^2 \b}{C})(1-\frac{\l^2 \b}{C})}
\mathrm{arctanh} \left[ \frac{(1-\frac{\l^2 \b}{C})}{(1+\frac{\l^2 \b}{C})}
\tanh \omega y \right] \, . \ee
 The new Schwarzschild-type metric is then
\be
\label{sigmetric} ds^2 = -U(\sigma) dt^2 + \frac{d\sigma^2}{U(\sigma)} \, ,
\ee where
\be
\label{sigu} U(\sigma) = \frac{1-\frac{(1+\frac{\l^2 \b}{C})^2}{(1-\frac{\l^2
\b}{C})^2} \tanh^2 \left( (1+\frac{\l^2 \b}{C})(1-\frac{\l^2 \b}{C}) \omega
\sigma \right)}{(1+\frac{\l^2 \b}{C})^2 \left[ 1 - \tanh^2 \left
(1+\frac{\l^2 \b}{C})(1-\frac{\l^2 \b}{C}) \omega \sigma \right) \right]} \,
. \ee The horizons $U(\sigma_{\pm})=0$ are
\be
\sigma_{\pm} = \pm \frac{\mathrm{arctanh} \frac{(1-\frac{\l^2 \b}{C})}
{(1+\frac{\l^2 \b}{C})}}{(1+\frac{\l^2 \b}{C})(1-\frac{\l^2 \b}{C}) \omega}
\, . \ee In these coordinates we can study the degenerate case $C=\l^2 \b$
since they will still be able to ``see'' the region between the horizons. In
this limit the solution becomes
\be
U(\sigma) = \frac{1-(4\omega\sigma)^2}{4} \,  \ee and the horizons still
remain uncoincident
\be
\sigma_{\pm} = \pm \frac{1}{4\omega} \, . \ee

To get the horizon temperature we should construct the Euclidean metric
setting $it=\tau$ and identifying $\tau$ with an appropriate period in order
to remove the singularities. But this is not so in this case because the
Killing vector cannot be normalized at infinity as in the standard
Schwarzschild case, due to the presence of the singularities. Bousso and
Hawking \cite{bh2} give the correct prescription. We need to find the point
$\sigma_g$ for which the orbit of the Killing vector coincides with the
geodesic going through $\sigma_g$. In such a point the effects of both black
holes attractions balance out exactly and an observer will need no
acceleration ($\Gamma^{\r}_{\mu\nu}=0$) to stay there, just like an observer
at infinity in the standard Schwarzschild case. A straightforward calculation
shows that this point is just where both horizons coincide in the degenerate
case ($r=0$), that is $\sigma_g =0$. With the adequate normalization the
horizon temperatures are given by \cite{bh2}
\be
T_{\pm} = \frac{1}{2\pi} \frac{1}{2\sqrt{U(\sigma_g)}} \left| \frac{\partial
U}{\partial \sigma} \right|_{\sigma_{\pm}} \, , \ee and then we get
\be
T_+ = T_- = \frac{1}{2\pi} (1+\frac{\l^2 \b}{C}) \sqrt{-C} \,  \ee and in
the $C \rightarrow \l^2 \b$ limit
\be
T_+ = T_- \rightarrow \frac{\sqrt{-\l^2 \b}}{\pi} \, . \ee Note that the horizon
temperatures are always coincident in either non-degenerate or degenerate
case in a different way from the 4D Schwarzschild-de Sitter case. We can then
complete the physical picture of this model, the two mirror black holes are
at the same temperature. This feature will have some important consequences
on the semiclassical theory as we will see later.\\

Finally we have to note that the transformation (\ref{strans-}) is performed
in the region between the horizons $x^+x^-<0$. We can realize a new
transformation in order to take into account the black hole interiors
$x^+x^->0$
\be
\label{strans+} \omega x^{\pm} = e^{\pm \omega y^{\pm}} \, . \ee In this case
the static metric is
\be
ds^2 = \frac{-dt^2 + dy^2}{(1-\frac{\l^2 \b}{C})^2 - \frac{4 \l^2 \b}{C}
\sinh^2 \omega y} \,  \ee and a further transformation
\be
\sigma = \frac{1}{\omega (1+\frac{\l^2 \b}{C})(1-\frac{\l^2 \b}{C})}
\mathrm{arctanh} \left[ \frac{(1-\frac{\l^2 \b}{C})}{(1+\frac{\l^2 \b}{C})}
\mathrm{cotanh} \omega y \right] \,  \ee brings the metric into the same
geometry (\ref{sigmetric}), (\ref{sigu}) so that there is no difference from
the last one as expected. \\

We now wish to comment briefly on the case $\b>0$ where the physical 
picture is completely different, i.e. the singularities are timelike and in the region 
between the horizons the Killing vector is spacelike . The Kruskal diagram is 
similar to
that of a point electric charge in 2+1 dimensions \cite{poch}.
Formally, the analysis of this section can be repeated step by step for this
solution as well. When the two horizons become degenerate there is again a Ginsparg-Perry 
type transformation connecting the constant (now negative) curvature, $\f=0$ solution
with (\ref{Smetric}), (\ref{lulo}), in much the same the way as it has
been done in \cite{nood}
for the 2d dilaton-Maxwell gravity.
  
\section{The 2D Birkhoff's theorem revisited}
Now we analyse the existence of dilaton-constant solutions 
in a more general context. This feature leads us to perform
a revision of the 2D Birkhoff's theorem. 
Under some assumptions one can ensure that the general solution is given, 
up to space-time diffemorphisms, by a one-parameter family of 
static metrics \cite{lmk}.
The parameter, related with the ADM mass, is diffeomorphism invariant and
classifies all of them. In particular, there exists a Schwarzschild gauge in
which the solution is manifestly static and the dilaton field is linear in
the space-like coordinate.\\
Considering the gravitational sector of (\ref{action2D})
\be
\label{gase}
S=\frac{1}{2\pi} \int d^2x \sqrt{-g} \left( R\phi + 4\l^2 V(\f) \right) \, ,
\ee this solution is written as (\ref{birk1}), (\ref{birk2}), where $M$,
given by (\ref{ADMmass}), is the diffeomorphism invariant parameter. We shall
show that there is also another type of solutions. For certain potentials
there is, in fact, another static solution providing a constant curvature
space with a constant dilaton field.
 The equations of motion (\ref{eq1}),
(\ref{eq2}) of the above functional action in a static gauge $\frac{\partial
\f}{\partial t}=0=\frac{\partial \r} {\partial t}$ (where $x^{\pm}=t \pm x$)
are \bea \label{st1} -\frac{d^2\r}{dx^2} + 2\l^2 e^{2\r} \frac{dV}{d\f} &=& 0
\, , \\ \label{st2} -\frac{d^2\f}{dx^2} + 4\l^2 e^{2\r} V &=& 0 \, . \eea If
$\frac{d\f}{dx} \neq 0$ the equation (\ref{st2}) admits a first integral
\be
\label{first} -\frac{d\f}{dx} +4\l^2 \int dx e^{2\r} V(\f) = 4\l M \, , \ee
where $M$ is an integration constant and, using the constraints, the equation (\ref{st1}) turns
into
\be
\l e^{2\r} = \frac{d\f}{dx} \, . \ee 
The equation (\ref{first}) gives the
conformal factor $e^{2\r}=4J(\f)-4M$ and in the Schwarzschild gauge, defined
by $dr=e^{2\r}dx$, we get finally the set (\ref{birk1}), (\ref{birk2}). This
is essentially the Birkhoff's theorem \cite{lmk}. Now we are going to
consider the $\frac{d\f}{dx} =0$ case, that is, dilaton-constant solutions.
\footnote{The existence of these kind of solutions was already noted in 
\cite{zf}.}
This kind of solutions $\f=\f_0$ can only exist for certain potentials
$V(\f)$ satisfying
\be
\label{condition} V(\f_0)=0 \, , \; \; \; \left. \frac{dV(\f)}{d\f}
\right|_{\f_0} \neq 0 \, , \ee so that the equation (\ref{st2}) is trivially
satisfied and (\ref{st1}) becomes
\be
\label{const} \frac{d^2\r}{dx^2} + \frac{R_0}{2} e^{2\r} = 0 \, , \ee where
\be
R_0 = -4\l^2 \left. \frac{dV}{d\f} \right|_{\f_0} = \mathrm{const} \, . \ee
Thus these solutions lead to constant curvature spacetimes. 
Making the
coordinate change $dr=e^{2\r}dx$ into the Schwarzschild gauge the equation
(\ref{const}) is easily integrated and the solution is written as \bea
\label{sbirk1} ds^2 &=& -(k-\frac{R_0}{2}r^2)dt^2 +
(k-\frac{R_0}{2}r^2)^{-1}dr^2 \, , \\ \label{sbirk2} \f &=& \f_0 =
\mathrm{const} \, , \eea
where $k$ is an integration constant. \\
Obviously both solutions (\ref{birk1}), (\ref{birk2}) and (\ref{sbirk1}),
(\ref{sbirk2}) are not diffeomorphism connected as it is manifested by the
scalar dilaton function. 
Note that this last dilaton-constant solution is not
available for a generic potential $V(\f)$ but only for those satisfying the
conditions (\ref{condition}). One example is the $\sinh \b \f$ potential
(\ref{sinh}); another one is provided in Appendix A starting from Einstein-Maxwell
gravity in 4D.
 In conformal gauge, in the special limit $C \rightarrow \l^2
\b$, we obtained the dilaton-constant ($\f=0$) solution (\ref{ds}) with $M=0$
and constant curvature $R=R_0=-8\l^2\b$. In a manifestly static gauge, it reads
\be
\label{confstat} ds^2=-(1+4\l^2\b r^2)dt^2+(1+4\l^2\b r^2)^{-1}dr^2 \, . \ee
But $\f=0$ is just the dilaton-constant solution for the $\sinh \b \f$
potential: $V(0)=0$, $\left. \frac{dV(\f)}{d\f} \right|_{\f_0} \neq 0$ and
moreover $J(0)=0$ so that the expression (\ref{ADMmass}) becomes identically
zero. The above solution coincides with (\ref{sbirk1}) (with $k=1$). 
Now we can complete
our understanding on the $C \rightarrow \l^2 \b$ limit of the solution (\ref{sinhmetric}),
(\ref{sinhdilaton}) in Kruskal gauge. The $C \neq \l^2 \b$ case
coincides, up to diffeomorphisms, with the $M\neq 0$ parametrized solution
(\ref{birk1}), (\ref{birk2}) and the $C = \l^2 \b$ case with the
unparametrized solution (\ref{sbirk1}), (\ref{sbirk2}). These solutions are
different and they cannot be diffeomorphism connected. The special case $M=0$
in (\ref{birk1}), (\ref{birk2}), which at first sight we could be tempted to
identify with (\ref{sbirk1}), (\ref{sbirk2}), is the horizon coincident case
and the region I of Fig. I is reduced to the point $r=0$ where $\f=0$ and $R=-8\l^2\b$.
The transformation (\ref{gp1}), (\ref{gp2}) connects both gauges in a similar
way to the Ginsparg-Perry one and this suggests that there is a deep relation
between the existence of constant dilaton solutions and horizon degeneration.
In fact this is what happens in general and we shall show this in the remaining part of
this section.\\

Let us consider again the general solution (\ref{birk1}), (\ref{birk2}) for a
general potential $V(\f)$ and introduce $U(r)=4J(r)-4M$ so that the horizons
are the roots of $U(r)$. In order to study models with horizon degeneration
we want $U(r)$ to have two or more roots. Although all roots are distinct we
can always fit a value $M_0$ of the parameter $M$ for which two neighbouring
roots become coincident in, say, $r_0$ which is then a double root of $U(r)$.
The `critical' value of $M$ is $M_0=J(r_0)$ and the dilaton function at
this point is $\f_0=\l r_0$. Now, since $r_0$ is an extremal of $U(r)$, we
get \bea 0 &=& \left. \frac{dU}{dr} \right|_{r_0} = 4\l V(\f_0) \, , \\ 0
&\neq & \left. \frac{d^2U}{dr^2} \right|_{r_0} = 4\l^2 \left. \frac{dV}{d\f}
\right|_{\f_0} \, , \eea which are just the conditions (\ref{condition}), and then
$\f=\f_0$ gives the constant dilaton solution (\ref{sbirk1}), (\ref{sbirk2}). It is
straightforward to check that the opposite is true as well: 
if $\f_0$ is a constant dilaton
solution, $r_0=\frac{\f_0}{\l}$ is a degenerate horizon for $M=M_0=J(\f_0)$.\\

Let us now perform a perturbation around the degenerate radius of coincident
horizons, as it happens in the limit $M \rightarrow M_0$ and $U(r) \rightarrow
0$ between the two horizons. We write
\be
M=M_0 - \frac{k}{4} \epsilon^2 \, , \ee 
where $\epsilon\ll 1$ and $k$ is a constant with the same sign 
as $R_0$. The degenerate case corresponds
to $\epsilon \rightarrow 0$. We introduce a new coordinate pair
$(\tilde{t},\tilde{r})$ defined by
\be
\label{gp2D}
t=\frac{\tilde{t}}{\epsilon} \, , \; \; \; \;
r=r_0 + \epsilon \tilde{r} \, . \ee Expanding the function $U(r)$ in
powers of $r-r_0$ we get
\be
U(r) = (k-\frac{R_0}{2}\tilde{r}^2) \epsilon^2 + O(\epsilon^3) \, ,
\ee which finally turns (\ref{birk1}) into \be ds^2 = -\left(
k-\frac{R_0}{2}\tilde{r}^2 + O(\epsilon)\right) d\tilde{t}^2 +
\left( k-\frac{R_0}{2}r^2+ O(\epsilon)\right)^{-1} d\tilde{r}^2 \,
. \ee This in the "near-horizon" limit $\epsilon \rightarrow 0$ becomes (\ref{sbirk1}).\\
We end by noting that for $R_0<0$, i.e. the solution has constant negative curvature,
$k$ is negative and redefining it as $k\equiv -m$ this is nothing but the 
AdS$_2$ black hole.

\section{Semiclassical theory and conclusions}
We shall now make some semiclassical considerations concerning the $\sinh
\beta\phi$ model. The Hawking radiation is determined by the usual expression
\be 
<T^f_{--}> =\frac{N}{12}\left[\partial^2_{-}\rho-(\partial_{-}\rho)^2
-t_{-}\right] 
\ee
and we now show why the choice $t_-=0$ in Kruskal coordinates is the most natural one. 
The privileged point $(r=0)$ in which the Killing vector must be
normalized corresponds in Kruskal coordinates with the curve
\be
x^+x^-={1\over  C} \>.\label{crit} \ee 
If we calculate $<T_{--}>$  we get
\be
<T^f_{--}>={N(x^+)^2\over48}\left[ {1\over 1+\lambda^2\beta x^+x^-} -{1\over
C x^+x^-+{\lambda^2\beta\over C}}\right]^2 \>. \ee This expression exactly
vanishes when evaluated over the points of the curve (\ref{crit}). The
interpretation is then that because the two black holes placed at infinity
have the same temperature there is a compensation between the
Hawking radiation coming from each black hole giving no net Hawking flux.
The same considerations apply if we interchange $-$ with $+$ in the previous
formulas and we have  $t_+=0$ as well. 
We can also wonder if it makes sense to choose `evaporating'
boundary conditions $t_-\neq t_+$. At the classical level and by virtue
of Birkhoff theorem the solutions are parametrized by a single constant $C$
forcing the two black holes to have the same mass and temperature. However
at the semiclassical level the Birkhoff theorem no longer applies and we
could try for instance to increase the mass of one of the black holes and to see
whether or not a new equilibrium state is reached. 
 Moreover,  if in view of a higher-dimensional interpretation we 
restrict 
to the case $\phi>0$ then the physical spacetime contains only one 
black hole
and it would seem natural to impose boundary
conditions different from the ones used above. These questions and the
related 
semiclassical dynamical evolutions will be studied elsewhere.
\\
It is interesting to comment that in the Jackiw-Teitelboim limit the
curvature singularities disappear and we get contant curvature AdS$_2$ black
holes (if $\beta >0$). 
AdS$_2$ black holes 
 have been claimed
not to emit Hawking radiation \cite{kim} (if a nontrivial dilaton is present,
however, this might not be true, see \cite{cami}),
which is exactly what happens in our
$\sinh\beta\phi$ model although there the no radiation can be understood by
the presence of the mirror black hole. Therefore intuitively the AdS$_2$
black hole inherits the no radiation property of the more general model they
arise in a certain limit. 
This is not the case of the
 exponential model in which black holes
 evaporate \cite{cfn}. In this model and with the boundary conditions $t_{\pm}=0$
 the solutions represent black holes in equilibrium with a thermal
bath.
 So the role of the mirror black hole is interchanged with the
existence of
 external radiation incoming onto the black hole.
\section*{Acknowledgements} 
This research has been partially supported by the DGICYT, Spain.
J.C. acknowledges the Generalitat Valenciana for an FPI fellowship. D.J. Navarro wishes
to thank the Spanish Ministry of Education and Culture for a fellowship and 
the Physics Department of Stanford University for hospitality.
A.F. is supported by an INFN fellowship.
We want to thank A. Mikovic for interesting comments.

\appendix
\section*{Appendix A \\
AdS$_2 \times $S$^2$ Geometry in Einstein-Maxwell Theory}
In this appendix we shall describe a way to generate the Robinson-Bertotti
(AdS$_2$ $\times$S$^2$)
geometry in Einstein-Maxwell gravity based on the possibility 
of constructing
constant-dilaton solutions explained in Section 5.
Let us start with the Einstein-Maxwell action
\be
I={1\over16\pi G^{(4)}}\int d^4x\sqrt{-g^{(4)}}\left(R^{(4)}
-\left(F^{(4)}\right)^2
\right)\>.
\ee
If we impose spherical symmetry on the gauge field and the metric
\be
ds_{(4)}^2=g_{\mu\nu}dx^{\mu}dx^{\nu}+{\phi^2\over{2\lambda^2}}d\Omega^2
\>,
\ee
where $x^{\mu}=(t,r)$, $d\Omega^2$ is the metric on the two-sphere and 
$\lambda^{-1}$
 is the Planck lenght ($\lambda^{-2}=G^{(4)}$), the
dimensionally reduced action functional is \cite{Thomi}
\be
\int d^2 x\sqrt{-g}\left[{1\over2}\left({\phi^2\over4} R+{1\over2}g^{\mu\nu}
\partial_{\mu}\phi\partial_{\nu}\phi+\lambda^2\right)-{1\over8}\phi^2F^{\mu\nu}
F_{\mu\nu}\right]\>.
\ee
After an appropriate reparametrization 
\be
{\phi^2\over4}\rightarrow \phi
\>,\ee
\be
g_{\mu\nu}\rightarrow g_{\mu\nu}(2\phi)^{-{1\over2}}\>,
\ee
the two-dimensional action takes 
the form \cite{lmk2}
\be
\int d^2x\sqrt{-g}\left[{1\over2}\left(\phi R+\lambda^2 V(\phi)\right)
-{1\over4}W(\phi)F^{\mu\nu}F_{\mu\nu}\right]\>,
\ee
where
\be
V(\phi)={1\over\sqrt{2\phi}},\qquad W(\phi)=(2\phi)^{3\over2}
\>.
\ee
The equations of motion imply that \cite{lmk2}
\be
F=q{e^{2\rho}\over W(\phi)}\>,
\ee
where $F=2F_{+-}$ and $q$ is a constant.
Substituting the above solution for $F$ into the other equations of motion one finds
that they are equivalent to those of the model
(\ref{gase}) with the replacement
\be
V(\phi)\rightarrow
V(\phi)-{q^2\over{\lambda^2 W(\phi)}}=V_{eff}\>
\ee
and so in our case
\be
V_{eff}={1\over\sqrt{2\phi}}-{q^2\over {\lambda^2(2\phi)^{3\over2}}}\>
\ee
and we can apply the arguments of section 5. We then have 
a constant dilaton solution $\phi=\phi_0$ for 
\be
V_{eff}=0\>
\ee
and therefore
\be
\phi_0={q^2\over{2 \lambda^2}}\>,
\ee
which turns out to be the radius of the horizon for the extremal 
Reissner-Nordstrom solution $r_+=r_-={1\over\lambda}\sqrt{2\phi_0}={q\over\lambda^2}$.
Moreover the two-dimensional
geometry is AdS$_2$ with curvature
\be
R=-{2\lambda^4\over q^2}=-{2\over r_+^2}\>.
\ee
%%%%%%%%%%%%%%%%%%%%%%%%%%%%%%%%%%%%%%%%%%%%%%%%%%%%%%%%%%%%%%%%%%%%%%%%%%%

\end{document}